# Classification of Microscopy Images of Breast Tissue: Region Duplication based Self-Supervision vs. Off-the Shelf Deep Representations

Aravind Ravi
Department of Systems Design Engineering
University of Waterloo, Canada
aravind.ravi@uwaterloo.ca

*Abstract*— Breast cancer is one of the leading causes of female mortality in the world. This can be reduced when diagnoses are performed at the early stages of progression. Further, the efficiency of the process can be significantly improved with computer aided diagnosis. Deep learning based approaches have been successfully applied to achieve this. One of the limiting factors for training deep networks in a supervised manner is the dependency on large amounts of expert annotated data. In reality, large amounts of unlabelled data and only small amounts of expert annotated data are available. In such scenarios, transfer learning approaches and self-supervised learning (SSL) based approaches can be leveraged. In this study, we propose a novel self-supervision pretext task to train a convolutional neural network (CNN) and extract domain specific features. This method was compared with deep features extracted using pre-trained CNNs such as DenseNet-121 and ResNet-50 trained on ImageNet. Additionally, two types of patch-combination methods were introduced and compared with majority voting. The methods were validated on the BACH microscopy images dataset. Results indicated that the best performance of 99% sensitivity was achieved for the deep features extracted using ResNet50 with concatenation of patch-level embedding. Preliminary results of SSL to extract domain specific features indicated that with just 15% of unlabelled data a high sensitivity of 94% can be achieved for a four class classification of microscopy images.

## I. Introduction

Among cancer-related death, breast cancer is one of the leading causes of death in women around the world. According to the Canadian Cancer Society, it is estimated that in 2020, 27,400 women will be diagnosed with breast cancer which represents 25% of all new cancer cases in women. Among the diagnosed cases of cancers leading to death in women, 13% are due to breast cancer [1]. A biopsy is the only definite way to diagnose breast cancer. During a biopsy, tissues samples are collected and examined by a pathologist to confirm the presence or absence of cancer cells in the sample.

Typically the tissue samples or microscopy slides are stained with the haematoxylin and eosin (H&E) before analysing the cells. The H&E staining enhances the appearance of the tissue with the cell nuclei appearing purple/dark blue in colour and cytoplasm as pink colour. Further it allows to distinguish tissues into one of the following types: normal, benign (non-malignant) and malignant. Lesions that represent changes in normal structures and do not spread to neighbouring areas are benign. Uncontrolled cell growth and spread to neighbouring areas are termed malignant and referred to as carcinomas. They can be further classified into in-situ (restricted to a local region) and invasive (invade/spread to neighbouring tissues). Finally, the pathologist provides a cancer diagnosis by analysing the cellular structure and organization of cells in the tissue sample. Digital images of tissue samples have gained increased application and usage in the development of computer aided diagnostic (CAD) tools to assist pathologists in cancer diagnosis.

Deep learning methods have been successfully applied to digital microscopy images and have shown success in automated diagnosis and classification of histopathology images [2]. Convolutional Neural Networks (CNNs) have been successfully applied to digital images for automatic feature extraction and classification. When large amounts of annotated data are available, supervised learning techniques can be applied. These methods can provide high performance and generalization in cases where large and diverse amounts of annotated data are available. In reality, large amounts of unlabelled data and only small amounts of annotated data are available. Therefore, in these scenarios, approaches such as transfer learning [3] and self-supervised learning (SSL) [4] can be leveraged to learn good features or representations.

In this study, four different types of feature extractors were compared on microscopy images of breast cancer with the goal of automatic classification. A novel self-supervised pretext task for domain specific feature extraction was proposed and compared with two other feature extractors based on transfer learning or pre-trained networks. Finally, these methods were applied on the BACH dataset with the objective of classifying microscopy images into one of four categories: normal, benign, in-situ and invasive. One common approach to classifying histopathology images is to segment a given image into sub-patches and pool the classification results of the sub-patches to arrive at a label for the original image, for example, by majority voting [5] [6]. In this study, deep features or embedding of sub-patches were extracted and two types of patch-combination methods based on concatenation and summation were introduced. These results were compared with the traditional majority voting based classifier.

The paper is organized as follows: Section II provides some of the related work and Section III introduces to the methods such as deep feature extraction, novel SSL pretext task, classification and performance evaluation used in this study. Section IV provides the results and discusses them. Finally, Section V concludes the paper with directions for future work.

II. RELATED WORK

Traditionally, computer-aided diagnosis systems for histopathology have been designed based on hand-crafted feature extraction methods. Some examples include analysis and segmentation of nuclei [7], analysing the morphological, structural and texture features, local-binary patterns (LBP) [8], etc. One of the major advantages of deep learning methods over hand-crafted features, is the capability of automatic feature extraction and classification. For the task of image classification and object detection, the convolutional neural network (CNN) based methods have outperformed traditional algorithms. CNNs can be used to learn good features and simultaneously be used for classification by optimizing a loss function. Many studies have reported success in applying CNNs for histopathology analysis and classification [2] [9]–[11].

One of the challenges in training deep learning models, is the need for large amounts of expert annotated data. In digital pathology acquiring large amounts of expert level annotated data is a challenge. Simultaneously, large amounts of unlabelled data is available. Therefore, this poses a unique opportunity to exploit domain adaptation methods such as transfer learning and unsupervised domain-specific feature extraction methods such as self-supervised learning [4][12].

*1) Transfer Learning:* Recently, there has been increased interest and success in applying pre-trained CNNs, trained on the ImageNet database [13], to various other types of images for feature extraction and training. The availability of pre-trained "off-the-shelf" open source CNNs has enabled for the accelerated development of methods in this area [14]–[16]. This technique of using a pre-trained network trained on a source domain (Ex: ImageNet) and extracting deep embeddings or fine-tuning the network on a target domain (Ex: BACH) is called transfer learning [5]. Since the earlier layers of CNNs learn generic image features such as edges, colour blobs and shapes, they can be used as a generic fixed feature extractor for images. One of the commonly used approaches for extracting deep embeddings is based on freezing the weights of the pre-trained CNN and use it as a fixed feature extractor. Next, the output of the last pooling layer are extracted on images from the target domain. This way, the generic image representations are considered to be transferred from the soure to the target domain without the need for lage amounts of annotated data. Another promising approach is based on domain adaptation of features by self-supervised learning.

*2) Self-supervised Learning:* When large amounts of unlabelled data and limited annotated data are available, the self-supervised learning approach can be used to extract domain specific features from the target dataset [17]. This method is similar to unsupervised learning, but the model is supervised based on "free" labels that come along with the data [18][19].

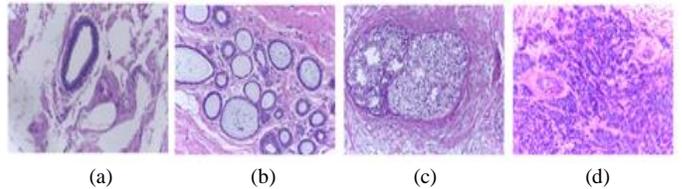

Fig. 1. Examples of microscopy images of size 2084 x 1536 from the BACH dataset. (a) normal, (b) benign, (c) in-situ carcinoma and (d) invasive carcinoma.

Typical self-supervised tasks comprise of a pretext task and a downstream task. In the pretext task, the model is trained based on the free labels or artifically generated labels. They can represent various contents of an image such as its geometry, orientation, rotation, or contextual content. For example, an artificial label can be generated for each image by rotating it in four different angles and train a four-class CNN classifier (pretext task) to predict the rotation angle of the image [20]. This method has been shown to extract meaningful image features that are subsequently used in a downstream task such as classification [21]. For example, if the downstream task is classification, then weights of the CNN from the pretext task, which are specific to the target domain, learned on a large set of unlabelled data are now used to train or finetune the CNN based on the limited set of annotated data. An alternative method is to use the embeddings extracted from the pretext task and apply a classifier such as Support Vector Machine (SVM) for the downstream task. Recently, self-supervised learning has been applied in medical images and for digital pathology to extract domain specific features [19]. A comprehensive review and comparison of self-supervised techniques applied in digital pathology is provided in [17]. The authors in [17] proposed an approach called the JigMag method which was based on a pretext puzzle task that solves a jigsaw magnification puzzle. A given image was converted to four different magnification levels and each magnification level was arranged in specific regions of a 2 x 2 grid, resulting in twelve different combinations of magnification and arrangements. Next, a CNN was trained with the objective of classifying a given 2 x 2 grid of JigMag into one of the twelve classes. The authors showed that the JigMag method provided the best domain specific features that enabled in highest classification performance. In this study, a novel self-supervised pretext task is porposed and this was inspired by the JigMag method.

III. METHODOLOGY

*A. BACH Dataset and Preprocessing*

All experiments and methods were developed and validated on the microscopy dataset introduced in the ICIAR 2018 grand challenge called the BreAst Cancer Histology (BACH) dataset [22]. This dataset comprises of 400 Haematoxylin and Eosin (H&E) stained microscopy images of breast tissue. It comprises of images belonging to four categories of breast tissue such as: normal tissue, benign tissue, in-situ carcinoma and invasive carcinoma. Every category of tissue type consists of 100 images with a high resolution of 2048 x 1536 pixels at 200x magnification and pixel size of 0.42μm x 0.42μm. Each image was a patch extracted from whole-slide images labelled by two pathologists. The dataset was downloaded from https://iciar2018-challenge.grand-challenge.org/dataset/. The goal of the challenge for microscopy images was to provide automatic classification of the tissue types. Figure 1 illustrates one example image for each category of breast tissue.

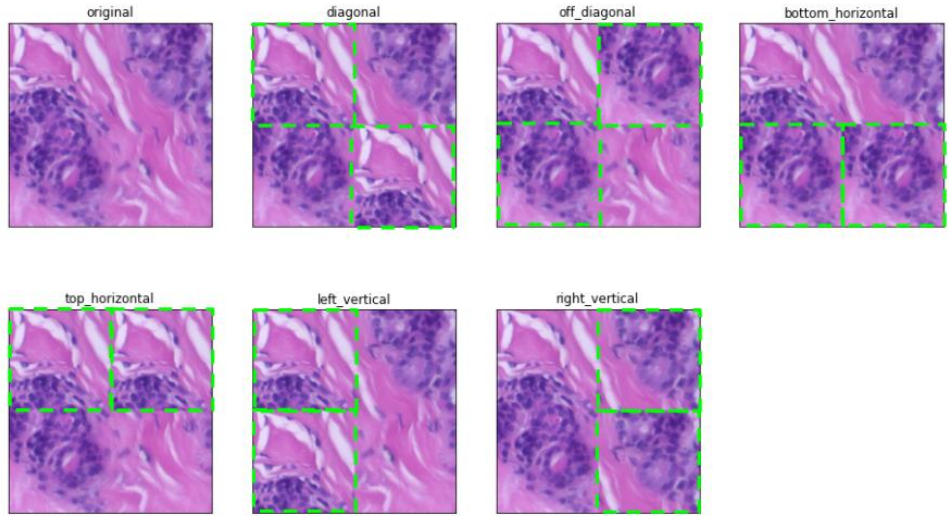

Fig. 2. Region duplication pretext task for self-supervised learning and feature extraction. The figure illustrates the six different ways of duplicating the regions for a single image patch of 512 x 512: diagonal, off-diagonal, bottom horizontal, top-horizontal, left-vertical and right-vertical.

Due to the small number of available images and large image dimensions, the images were segmented into smaller sub-patches of shape 512 x 512 pixels without overlap similar to [23]. As a result, each image was segmented into 12 patches and each patch was assigned the same label as the combined image-wise label. Finally, a total of 4800 raw patches were used to develop the methods. These raw patches were provided as input to all the methods discussed in subsequent sections and no additional pre-processing was performed.

### B. Representation Learning Tasks

Two types of features or learned representations were compared. The first method was based on deep features or deep embedding extracted using pre-trained CNNs referred to as transfer learning. The second method was based on a novel self-supervised learning pretext task.

*1) Transfer Learning:* Two different pre-trained networks were used as fixed feature extractors: DenseNet-121 [24] and ResNet50 [25]. These were pre-trained networks trained on ImageNet database. For the microcopy images used in this study, each segmented image patch of size 512 x 512 was provided as input to the pre-trained network and the output of the last pooling layer was extracted. For a single image patch, this resulted in an embedding vector of length 1024 and 2048 for the DenseNet-121 and ResNet50 respectively.

*2) Self-supervised Learning:* The DenseNet-121 architecture initialized with random weights was used to train on a novel pretext task to extract domain specific features. A novel task called region duplication identification is proposed to extract domain specific features for the BACH dataset.

### C. Region Duplication – Pretext Task

Microscopy images resemble textures and have repeting patterns, therefore, a pretext task such as the region duplication detection was used. For solving this task, the network should focus on the regions of the image that have high similarity or duplication. In the process, the CNN learns to extract meaningful features or repetitive patterns from histology images. To generate the "free" labels from the unlabelled data, the 512 x 512 image patch was divided into a 2 x 2 grid. Consider the image patch shown in Figure 2 (top left). In this 2 x 2 grid, there are 6 ways of duplication possible: diagonal, off-diagonal, bottom-horizontal, top-horizontal, left-vertical and right-vertical. For example, duplicating a region means, the sub-patch in the top-left grid was copied to the top right grid and replaced in place of the previous set of pixels. Similarly, the top-left grid can be duplicated along the diagonal and copied into the bottom right grid. By duplicating the regions, for a single patch, 6 artificially synthesized images were generated.

The DenseNet-121 (DNet) architecture was selected for the pretext task as there are less number of trainable parameters compared to the ResNet50 architecture. The DNet was initialized with random weights before training the CNN on the task of region duplication classification. The CNN was trained to classify the images into one of seven classes of duplication: normal (original patch), top horizontal, bottom horizontal, right vertical, left vertical, diagonal and off-diagonal.

The CNN was initialzed with random weights and trained to minimize the cross-entropy loss function. The model was trained under two conditions: 10% and 15% of the unlabelled training data. For each condition, a given patch was converted into seven patches, each belonging to one of the duplication strategies. As a result, for the 10% case, 40 images were converted into 480 sub-patches and 3360 new sub-patches. Similarly for the 15% case, 60 images were converted to 720 sub-patches and 5040 new sub-patches. The new sub-patches were provided as input to the model and was trained on the proposed SSL pretext task. The trained models are referred to as S-DNet-10 and S-DNet-15 respectively. All models were developed in Python using the Keras Tensorflow API and trained on the GPU provided by the Google colaboratory environment with 12 GB RAM. The following were the training parameters of the CNN: batch size = 16, learning rate = 0.0001, number of epochs = 60 and the catergorical cross-entropy loss function.

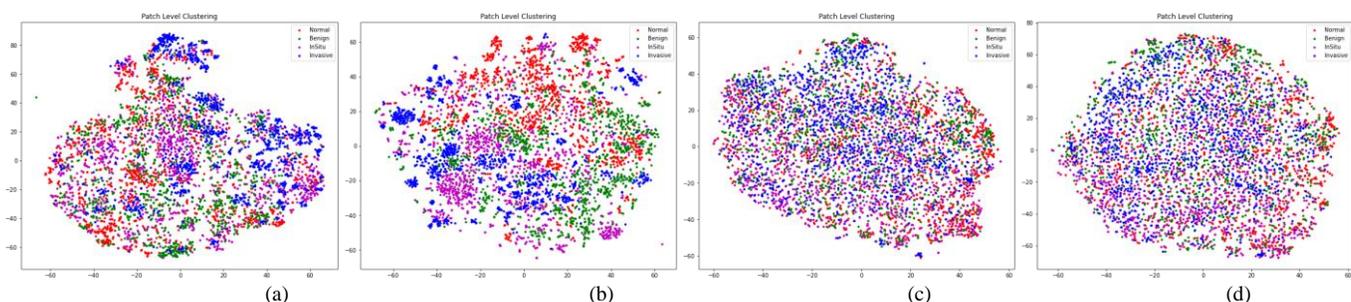

Fig. 3. t-SNE of patch-wise features extracted using all four feature extractors: (a) P-DNet, (b) P-RNet, (c) S-DNet-10 and (d) S-DNet-15. The clusters are colored according to the respective class labels: normal (red), benign (green), in-situ (purple) and invasive (blue).

### D. Classification

Features extracted based on pre-trained networks such as DenseNet-121 and ResNet50 on microscopy images of breast cancer were compared with the features extracted using the models trained based on the proposed SSL task. As a result, four feature extractors were compared: pre-trained DenseNet-121 (P-DNet), pre-trained ResNet50 (P-RNet), self-supervised DenseNet-121 at 10% of unlabelled examples (S-DNet-10) and self-supervised DenseNet-121 at 15% of unlabelled examples (S-DNet-15). The deep embedding for each patch of size 512 x 512 was extracted from the last pooling layer of all four feature extractors. Classifiers were developed to classify the image patches into one of four categories: normal, benign, in-situ and invasive. Classification was performed at two levels: at the patch-level (images of size 512 x 512) and at the slice or image-level.

*1) Patch-wise classification:* For the patch-level classifier a support vector machine (SVM) with a radial basis function was trained. The following parameters were selected: regularization parameter C=10 and gamma=0.001.

*2) Slice-level classification:* For the image or slice-level classifier, three different methods were compared: (i) a majority voting technique, (ii) concatenation of patch-wise deep embedding and (iii) summation of patch-wise deep embedding. Concatenation of patch-wise embedding: embedding of the 12 sub-patches were concatenated into one long feature vector. For example: the patch-wise feature vector length for DenseNet-121 was 1024 and as a result of concatenating the patch-level embedding the final feature vector length for the combined slice was 12288. The same procedure was performed for all types of feature extractors and all images. The order in which the patches were combined was kept consistent for all images and methods.

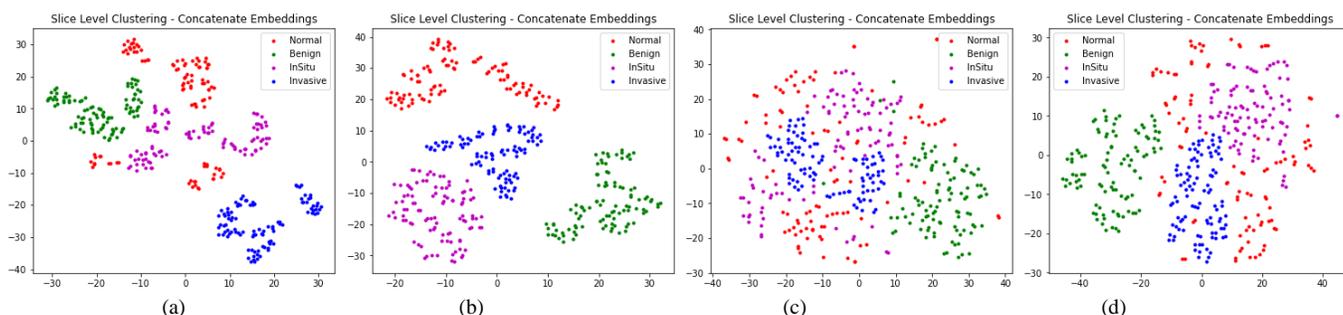

Fig. 4. t-SNE of slice-level features extracted by concatenating the patch-wise deep embedding using all four feature extractors: (a) P-DNet, (b) P-RNet, (c) S-DNet-10 and (d) S-DNet-15. The clusters are colored according to the respective class labels: normal (red), benign (green), in-situ (purple) and invasive (blue).

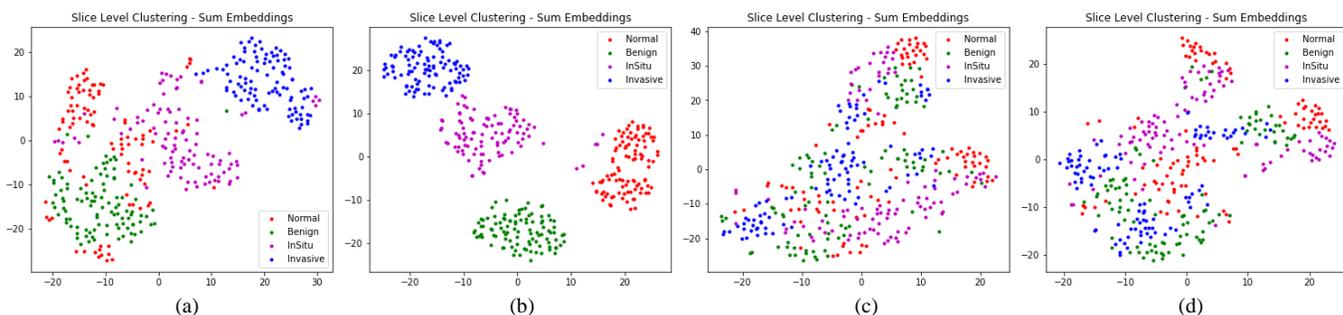

Fig. 5. t-SNE of slice-level features extracted by summation of patch-wise deep embedding using all four feature extractors: (a) P-DNet, (b) P-RNet, (c) S-DNet-10 and (d) S-DNet-15. The clusters are colored according to the respective class labels: normal (red), benign (green), in-situ (purple) and invasive (blue).

Summation of patch-wise embedding: this method was simply the sum of all the 12 patch-wise embedding, resulting in a single vector representing the entire slice. A linear four class SVM classifier with regularization parameter C=10 was used for both types of patch-level embedding combination methods.

*E. Performance Evaluation*

The objective of this study was to extract good features or representations for microscopy images of breast tissue and subsequently use them to automatically classify into one of four categories. To evaluate the learned representations of the four different feature extractors, the t-Stochastic Neighborhood Embedding (t-SNE) method was used [26]. The t-SNE allows to visualize a high dimensional vector in two dimensions. Typically, good features lead to reasonable clustering that enables to distinguish between different categories of data. Therefore, t-SNE was applied to visualize the features for patch-level clustering and slice-level clustering.

Next, the classification performance was measured based on the sensitivity score for each feature extraction method and classification algorithm. For the patch-level classification, the dataset was split into 75% training data and 25% test data. For the slice-level classification, the majority voting based technique was applied wherein the patch-level classifier was used to predict the label for a given patch and finally the class that was predicted the most number of times out of the 12 individual patches was assigned to the combined slice. These results were reported on the 25% test data. Finally, for both the concatenation and summation of embedding, a 4-fold cross-validation that comprised of 75% training data and 25% test data was used to measure the performance. For all cases the overall sensitivity and class-wise sensitivity are reported.

## IV. RESULTS AND DISCUSSIONS

*A. Deep Embeddings Visualization*

The learned representations or deep embedding for all four feature extractors were visualized by applying t-SNE and reducing the dimensions to 2D. The four feature extractors were: the pre-trained DenseNet-121 (P-DNet), pre-trained ResNet50 (P-RNet), self-supervised DenseNet-121 with 10% data (S-DNet-10) and self-supervised DenseNet-121 with 15% data (S-DNet-15). Figure 3 illustrates the patch-wise clustering for all four feature extractors. Among the pre-trained and self-supervised features, a clear clustering of patch embedding can be observed for features extracted using the pre-trained networks. Moreover, the features extracted using the pre-trained ResNet50 CNN provides the best clustering. Among the self-supervised methods, clustering can be observed for the patches belonging to the normal and invasive categories. Between the S-DNet-10 and S-DNet-15, S-DNet-15 provides comparatively better clustering. This can be attributed to the increase in the unlabelled training data size from 10% to 15%.

Next, Figures 4 and 5 illustrate the t-SNE visualizations of features extracted for slice-level clustering based on the patch-combination methods. Across all feature extractors, distinct clusters of the slices/images can be observed when the deep embedding were concatenated. Moreover, for the self-supervised features, even though there was less separation among clusters at the patch-level, the slice-level clusters illustrate good clustering and class separation. The results indicate that the feature extractors provide good representations for the images, and could potentially lead to improved classification performance at the slice-level.

TABLE I. PATCH-WISE CLASSIFICATION PERFORMANCE (SENSITIVITY) ACROSS ALL FEATURE EXTRACTORS

| FEATURES | SENSITIVITY (%) | | | | |
|---|---|---|---|---|---|
| | NORMAL | BENIGN | IN-SITU | INVASIVE | OVERALL |
| S-DNET-10 | 73.3 | 58.6 | 58.3 | 58.6 | **62.3** |
| S-DNET-15 | 74 | 52.6 | 54.6 | 51 | **58.1** |
| P-DNET | 65.3 | 66.6 | 61.3 | 63 | **64.1** |
| P-RNET | 89 | 72.6 | 72.6 | 64 | **74.6** |

*B. Breast Cancer Classification*

The classification performance was measured based on the sensitivity or accuracy metric. The performance was measured for patch-wise classification and slice-level or image-level classification. Both class-wise and overall sensitivity scores were calculated.

Table I presents the overall and class-wise classification results for all four feature extractors when the classifier was applied at the patch-level. Among the four compared feature extractors, the highest overall sensitivity of 74.6% was achieved for the features extracted using the P-RNet. Among the self-supervised methods, the S-DNet-10 provides an overall patch-wise sensitivity of 62.3%. The patch-wise results are of particular interest as they will directly influence the slice-level

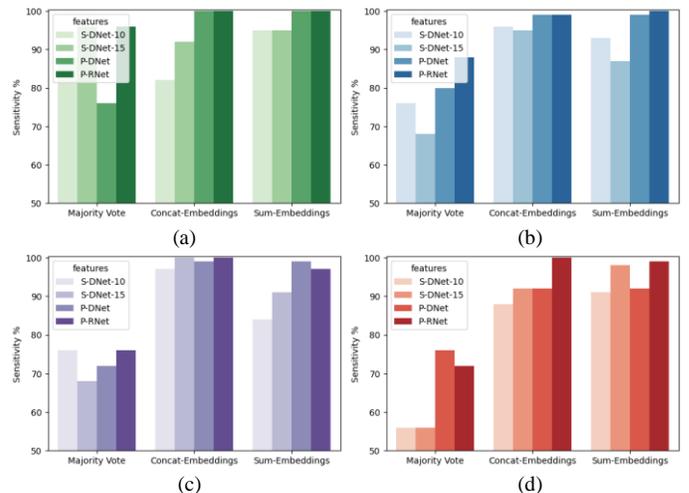

Fig. 6. Slice-level classification performance (sensitivity) for each tissue type: normal (a), benign (b), in-situ carcinoma (c) and invasive carcinoma (d) across all feature extractors and patch-combination methods.

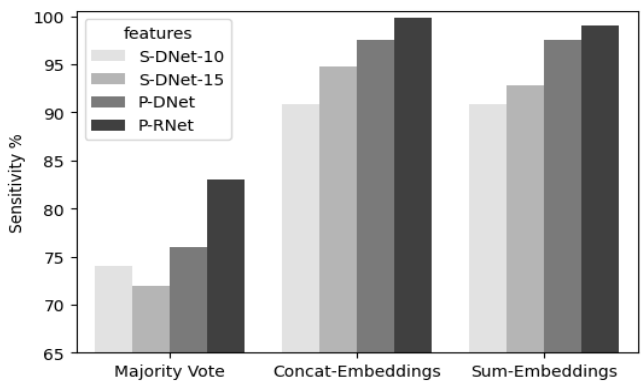

Fig. 7. Overall slice-level classification performance (sensitivity) across all feature extractors and patch-combination methods.

or image-level classification, especially for patch-combination methods such as majority voting.

Next, the slice-level or image-level classification performances were compared for three different types of patch-combination methods: majority voting, concatenating the deep embedding and summation of the deep embedding. The results are summarized in Figures 6 and 7. Figure 6, illustrates the slice-level sensitivity for each tissue category: normal, benign, in-situ carcinoma and invasive carcinoma. It can be observed that for all tissue types, the best performing method is based on concatenating the deep embedding. Both concatenation and summation of the deep embedding outperforms the standard majority voting based method.

Figure 7 illustrates the overall sensitivity of the slice-level or image-level classifier. An overall sensitivity of 99% was achieved when the patch-wise features of P-RNet were concatenated compared to the summation approach (97%) or majority voting (84%). It is interesting to note the high performance of the features extracted using the self-supervised methods. Both S-DNet-10 and S-DNet-15 features when concatenated, provide high classification sensitivity of 90% and 94% respectively. Among the three DNets, despite the low number of unlabelled training examples, 10% and 15% of training examples, the D-Nets generally provide good features such that when concatenated, provides high classification sensitivity at the slice-level.

This result of the SSL methods are of particular interest, because with a very small number of unlabelled and domain specific images (10% and 15%) the proposed pretext task provides good features leading to high performance at the slice or image-level. Compared to the pre-trained networks which were trained on approximately over 1 million non-medical images belonging to 1000 categories, the self-supervised methods can provide both domain specific features and high performance at a much lower annotation budget. Further investigation will be required to determine the influence of unlabelled training data size on the performance of the downstream task.

## V. CONCLUSIONS AND FUTURE WORK

In this study, a novel SSL pretext task for microscopy images of breast tissue (BACH dataset) was proposed. The task of region duplication identification was designed to extract domain specific features. Deep features extracted using pre-trained DenseNet-121 and ResNet-50 were used a baseline for comparison. Furthermore, two types of patch-combination methods based on concatenating and summation of patch-level embedding were introduced and compared with majority voting. The results indicated that an overall cross-validated performance of 99% sensitivity was achieved for deep features of pre-trained ResNet50 with concatenation of patch-level embedding. Finally, preliminary results of the proposed SSL domain specific features indicate that with just 15% of unlabelled data, a high sensitivity of 94% can be achieved for a four class classification of microscopy images.

Some limitations of this study are discussed and will be studied further to be addressed in future studies. The SVM classifier was directly applied on the deep features of the SSL model, whereas improvement in performance can be achieved if the model was fine-tuned with labelled data. The proposed approaches were only validated on the BACH dataset, therefore, future studies can explore the applicability of the methods on other digital pathology datasets. Finally, the amount of unlabelled data used in this study was 10% and 15%, therefore, further investigation on larger unlabelled datasets are required to validate the proposed self-supervised approach. Overall, the preliminary results of the proposed self-supervised task shows promising ability of the network to learn domain specific features for breast cancer classification of histopathology images.

REFERENCES

[1] "https://www.cancer.ca/en/cancer-information/cancer-type/breast/statistics/?region=on (accessed: Dec. 17, 2020).".

[2] C. L. Srinidhi, O. Ciga, and A. L. Martel, "Deep neural network models for computational histopathology: A survey," *Med. Image Anal.*, vol. 67, p. 101813, 2020.

[3] M. Kohl, C. Walz, F. Ludwig, S. Braunewell, and M. Baust, "Assessment of Breast Cancer Histology Using Densely Connected Convolutional Networks," *Lect. Notes Comput. Sci. (including Subser. Lect. Notes Artif. Intell. Lect. Notes Bioinformatics)*, vol. 10882 LNCS, pp. 903–913, 2018.

[4] L. Jing and Y. Tian, "Self-supervised Visual Feature Learning with Deep Neural Networks: A Survey," *IEEE Trans. Pattern Anal. Mach. Intell.*, vol. 8828, no. c, pp. 1–1, 2020.

[5] L. Alzubaidi, O. Al-Shamma, M. A. Fadhel, L. Farhan, J. Zhang, and Y. Duan, "Optimizing the performance of breast cancer classification by employing the same domain transfer learning from hybrid deep convolutional neural network model," *Electron.*, vol. 9, no. 3, 2020.

[6] T. S. Sheikh, Y. Lee, and M. Cho, "Histopathological classification of breast cancer images using a multi-scale input and multi-feature network," *Cancers (Basel).*, vol. 12, no. 8, pp. 1–21, 2020.

[7] Y. M. George, H. H. Zayed, M. I. Roushdy, and B. M.

Elbagoury, "Remote computer-aided breast cancer detection and diagnosis system based on cytological images," *IEEE Syst. J.*, vol. 8, no. 3, pp. 949–964, 2014.

[8] M. D. Kumar, M. Babaie, S. Zhu, S. Kalra, and H. R. Tizhoosh, "A Comparative Study of CNN, BoVW and LBP for Classification of Histopathological Images," *arXiv*, 2017.

[9] T. Araujo *et al.*, "Classification of breast cancer histology images using convolutional neural networks," *PLoS One*, vol. 12, no. 6, pp. 1–14, 2017.

[10] Y. Li, X. Xie, L. Shen, and S. Liu, "Reversed active learning based atrous densenet for pathological image classification," *arXiv*, 2018.

[11] A. Rakhlin, A. Shvets, V. Iglovikov, and A. A. Kalinin, "Deep Convolutional Neural Networks for Breast Cancer Histology Image Analysis," *Image Anal. Recognition. ICIAR 2018. Lect. Notes Comput. Sci.*, vol. 10882, 2018.

[12] M. Sikaroudi, A. Safarpoor, B. Ghojogh, S. Shafiei, M. Crowley, and H. R. Tizhoosh, "Supervision and Source Domain Impact on Representation Learning: A Histopathology Case Study," *Proc. Annu. Int. Conf. IEEE Eng. Med. Biol. Soc. EMBS*, vol. 2020–July, pp. 1400–1403, 2020.

[13] A. Krizhevsky, I. Sutskever, and G. E. Hinton, "ImageNet Classification with Deep Convolutional Neural Networks," in *Advances in Neural Information Processing Systems*, 2012, pp. 1097–1105.

[14] A. S. Razavin, H. Azizpour, J. Sullivan, and S. Carlsson, "CNN Features off-the-shelf: an Astounding Baseline for Recognition," in *Proceedings of the IEEE Computer Society Conference on Computer Vision and Pattern Recognition*, 2014, pp. 806–813.

[15] B. Kieffer, M. Babaie, S. Kalra, and H. R. Tizhoosh, "Convolutional neural networks for histopathology image classification: Training vs. Using pre-trained networks," *Proc. 7th Int. Conf. Image Process. Theory, Tools Appl. IPTA 2017*, vol. 2018–January, pp. 1–6, 2018.

[16] C. Szegedy, V. Vanhoucke, S. Ioffe, J. Shlens, and Z. Wojna, "Rethinking the Inception Architecture for Computer Vision," *Proc. IEEE Comput. Soc. Conf. Comput. Vis. Pattern Recognit.*, vol. 2016–December, pp. 2818–2826, 2016.

[17] N. A. Koohbanani, B. Unnikrishnan, S. A. Khurram, P. Krishnaswamy, and N. Rajpoot, "Self-Path: Self-supervision for Classification of Pathology Images with Limited Annotations."

[18] V. Cheplygina, M. de Bruijne, and J. P. W. Pluim, "Not-so-supervised: A survey of semi-supervised, multi-instance, and transfer learning in medical image analysis," *Med. Image Anal.*, vol. 54, pp. 280–296, 2019.

[19] J. Gildenblat and E. Klaiman, "Self-Supervised Similarity Learning for Digital Pathology," pp. 1–9, 2019.

[20] S. Gidaris, P. Singh, and N. Komodakis, "Uneupervised Representation Learning by Predicting Image Rotations," in *International conference on learning representations*, 2018.

[21] L. Jing and Y. Tian, "Self-supervised Visual Feature Learning with Deep Neural Networks: A Survey," *IEEE Trans. Pattern Anal. Mach. Intell.*, pp. 1–1, 2020.

[22] G. Aresta and T. Ara, "BACH : grand challenge on breast cancer histology images," pp. 1–4, 2019.

[23] K. Nazeri, A. Aminpour, and M. Ebrahimi, "Two-Stage Convolutional Neural Network for Breast Cancer Histology Image Classification," *Lect. Notes Comput. Sci. (including Subser. Lect. Notes Artif. Intell. Lect. Notes Bioinformatics)*, vol. 10882 LNCS, pp. 717–726, 2018.

[24] G. Huang, Z. Liu, L. Van Der Maaten, and K. Q. Weinberger, "Densely connected convolutional networks," *Proc. - 30th IEEE Conf. Comput. Vis. Pattern Recognition, CVPR 2017*, vol. 2017–January, pp. 2261–2269, 2017.

[25] K. He, X. Zhang, S. Ren, and J. Sun, "Deep residual learning for image recognition," *Proc. IEEE Comput. Soc. Conf. Comput. Vis. Pattern Recognit.*, vol. 2016–December, pp. 770–778, 2016.

[26] G. H. Laurens van der Maaten, "Visualizing Data using t-SNE," *J. Mach. Learn. Res.*, vol. 9, pp. 2579–2605, 2008.